
\input phyzzx

\def\np{Nucl. Phys.}
\def\pl{Phys. Lett.}

\def\cmp{Comm. Math. Phys.}

\def\mpl{Mod. Phys. Lett.}

\def\phyrep{Phys. Rep.}

\tolerance=500000
\overfullrule=0pt
\Pubnum={US-FT-7/93}
\pubnum={US-FT-7/93}
\date={July, 1993}
\pubtype={}
\titlepage

\title{Topological current algebras in two dimensions}
\author{J. M. Isidro and A.V. Ramallo}
\address{Departamento de F\'\i sica de
Part\'\i culas \break Universidad de Santiago \break
E-15706 Santiago de Compostela, Spain}

\abstract{Two-dimensional topological field theories possessing a
non-abelian  current symmetry are constructed. The topological conformal
algebra of these models  is analysed. It  differs from the one obtained by
twisting the $N=2$ superconformal models and contains generators of
dimensions $1$, $2$ and $3$ that close a linear algebra.
Our construction can be carried out with one and two bosonic currents
and the resulting theories can be interpreted as topological
sigma models for  group manifolds.}

\endpage
\pagenumber=1
\sequentialequations

The study of two-dimensional topological conformal field theories
\REF\wittop{E. Witten \journal\cmp&117(88)353.}
\REF\bbrt{For a review see D. Birmingham, M. Blau, M.Rakowski
and G. Thompson \journal\phyrep&209(91)129.}[\wittop,\bbrt] has
attracted much interest due to its connections with
non-critical string
theories
\REF\witgrav{E. Witten \journal\np&B340(90)281.}
\REF\dij{R. Dijkgraaf, E. Verlinde and H. Verlinde
\journal\np&B352(91)59.;``Notes on topological string theory
and 2d quantum gravity", Proceedings of the Trieste spring
school 1990, edited by M.Green et al. (World Scientific,
Singapore,1991).}[\witgrav, \dij ]. In particular from these studies one
would like to get some insight  that might help to go beyond
the $d=1$ barrier in string theory
\REF\gins{For a review see P. Ginsparg and G. Moore,
``Lectures on 2D Gravity and  2D String
Theory"(hep-th/9304011), Yale preprint YCTP-P23-92.}[\gins]. A possible way
to attack this problem is by considering topological theories in which
additional symmetries are present. The simplest types of such theories are
the ones that enjoy a current algebra symmetry. The  obvious question raised
in this respect is whether or not a current algebra symmetry can coexist with
a topological conformal algebra. Let us recall a similar situation in
ordinary conformal field theories. In this case  it is well known that  the
Sugawara construction allows us to associate to any current algebra an
energy-momentum tensor such that the currents are dimension-one operators
with respect to it.

Our goal in this paper will be to find a
topological symmetry in which all generators can be represented in terms of
currents in a Sugawara-like form. We shall find that this symmetry is
generated by operators of dimensions $1$, $2$ and $3$, that are grouped in
topological doublets. In this point our algebra is different from the one
 obtained by twisting
\REF\EY{T. Eguchi and S.-K. Yang \journal\mpl&A4(90)1653;
T. Eguchi, S. Hosono and S.-K. Yang \journal\cmp&140(91)159.}[\EY] the $N=2$
superconformal models
\REF\LVW{W. Lerche, C. Vafa and N.P. Warner
\journal\np&B324(89)427.}[\LVW], in which
only operators of dimensions $1$ and $2$ appear. The presence of
dimension-three operators is an unavoidable consequence of having a
{\bf non-abelian } current algebra in our theory. We have found this type
of topological symmetry in our study of the $gl(N,N)$ affine
Lie superalgebras
\REF\nos{J. M. Isidro and A. V. Ramallo, `` $gl(N,N)$ current algebras and
topological field theories"(hep-th/9307037), Santiago preprint
US-FT-3/93.}[\nos]. Here we study this problem from a general point of view.

Let us consider a  Lie algebra $g$  generated by the
hermitian matrices $T^a$ \break $(a=1,\cdots, {\hbox{\rm dim}}\,g$)  that
satisfy the commutation relations
$$
[T^a,T^b]=if^{abc}T^c.
\eqn\uno
$$
The Lie algebra $g$ is assumed to be semisimple and the generators $T^a$ are
chosen in such a way that  $Tr(T^aT^b)=\delta_{ab}$. An holomorphic current
taking values on $g$ is an operator $J_a(z)$ whose Laurent modes are defined
 by
$$
J_a(z)=\sum_{n\in Z}J_a^n z^{-n-1}.
\eqn\dos
$$
The $g$-current algebra is obtained by requiring the modes $J_a^n$ to
satisfy the following commutation relations
$$
[J_a^n,J_b^m]=if^{abc}J_c^{n+m}
+kn\delta_{ab}\delta_{n+m,0}.
\eqn\tres
$$
In eq. \tres\ $k$ is a $c$-number constant (the level of the algebra). An
alternative definition of the current algebra is obtained by giving the
operator product expansion (OPE) of two currents
$$
J_a(z_1)J_b(z_2)={k\delta_{ab}\over
(z_1-z_2)^2}+if^{abc}{J_c(z_2)\over z_1-z_2}.
\eqn\cuatro
$$
The Sugawara energy-momentum tensor is given by a
bilinear expression in the currents. In order to unambiguously define
operators  in which two or more fields evaluated
at the same point are multiplied, we need to adopt a normal
ordering prescription. Suppose that $A(z)$ and $B(z)$ are two
local fields whose Laurent modes are $A^n$ and $B^n$,
$$
A(z)=\sum_{n\in Z}A^n z^{-n-\Delta_A}
\,\,\,\,\,\,\,\,\,\,\,\,\,\,
B(z)=\sum_{n\in Z}B^n z^{-n-\Delta_B},
\eqn\cinco
$$
where $\Delta_A$ and $\Delta_B$ are the conformal weights
of $A(z)$ and $B(z)$ respectively. All the fields we shall
encounter inside normal-ordered products will have integer
conformal weights and, therefore, we shall assume that this
condition is satisfied in the equations that follow.
The normal-ordered
product of two arbitrary modes $:A^n B^m:$ is defined as
$$
 :A^n B^m:\,\,\equiv
\cases{A^n B^m, &if $m \geq 1-\Delta_B $\cr
       (-1)^{g(A)g(B)} B^m A^n   , &if $m < 1-\Delta_B$,\cr}
\eqn\seis
$$
where $g(A)$ is equal to zero (one) if the operator $A$ is bosonic
(fermionic). The modes $(:AB:)^n$ of the normal-ordered product
of $A$ and $B$ are defined by the equation

$$
:A(z)B(z): \,\, \equiv
\sum_{n\in Z} (:AB:)^n z^{-n-\Delta_A -\Delta_B}.
\eqn\siete
$$
Substituting the mode expansions of $A(z)$ and $B(z)$ in
the left-hand side of \siete\ and using \seis\ we get
$$
(:AB:)^n= \sum_{p=1-\Delta_B}^{\infty} A^{n-p} B^p +
 (-1)^{g(A)g(B)}\sum_{p=\Delta_B}^{\infty}B^{-p}A^{n+p}.
\eqn\ocho
$$
The Sugawara energy-momentum tensor for the currents $J_a$ is given by
\REF\PW{A.M. Polyakov and P.B. Wiegmann \journal\pl&131(83)121.}
\REF\WZW{E. Witten \journal\cmp&92(84)455.}
\REF\KZ{V. G. Knizhnik and A. B. Zamolodchikov \journal\np&B247(84)83.}
[\PW,\WZW,\KZ]
$$
T(z)={1\over 2k+C_A}:J_a(z)J_a(z):
\eqn\nueve
$$
where $C_A$ is the quadratic Casimir of the adjoint representation of $g$:
$$
f^{acd}f^{bcd}=C_A\delta_{ab}.
\eqn\diez
$$
For simply-laced algebras $C_A$ is given by
$$
C_A={\bf \theta^2}({{\hbox{\rm dim}}g\over r}-1),
\eqn\once
$$
where $r$ is the rank of $g$ and ${\bf \theta }^2$
is the square of the lenght
of the  highest root of $g$.
For an $su(N)$ algebra, our conventions for the normalization of the $T^a$
matrices correspond to ${\bf \theta ^2}=2$ and, therefore, eq. \once\ gives
$C_A=2N$. It can be checked using standard methods that
$$
\eqalign{
T(z_1)J_a(z_2)=&{J_a(z_2)\over (z_1-z_2)^2}+
{\partial J_a(z_2)\over z_1-z_2}\cr
T(z_1)T(z_2)=&{c\over 2(z_1-z_2)^4}+
{2T(z_2)\over (z_1-z_2)^2}+
{\partial T(z_2)\over z_1-z_2},\cr}
\eqn\doce
$$
where the central charge $c$ in eq. \doce\ is given by [\KZ]
$$
c={2k\,{\hbox{\rm dim}}\,g\over 2k+C_A}.
\eqn\trece
$$
When dealing with normal-ordered products as those appearing in eq. \nueve,
one has to bear in mind that in general they are non-commutative and
non-associative. For example suppose that two fields $A$ and $B$, whose
Laurent models are defined as in eq. \cinco, satisfy
$$
[A^n,B^m]=D^{n+m}+nk_D\delta_{n+m,0},
\eqn\catorce
$$
where $k_D$ is a constant and the brackets denote a commutator if some of
the fields is bosonic and an anticommutator if both are fermionic. Then
using eq. \ocho\ one can easily prove that
$$
(:AB:)^n=(-1)^{g(A)g(B)}(:BA:)^n-(\partial D)^n
-{\Delta_D\over 2}(\Delta_A-\Delta_B)k_D\delta_{n,0},
\eqn\quince
$$
where $(\partial D)^n=-(n+\Delta_D)D^n$.
Unless otherwise specified, when more than two fields are multiplied the
normal order is defined inductively according to
the rule
$$
:A_n\cdots A_1:\equiv
:(:(:\cdots (:A_nA_{n-1}:)\cdots :)A_2 :)A_1):,
\eqn\dseis
$$
\ie, the product $:A_n\cdots A_1:$ is considered as the
product of $:A_n\cdots A_2:$ with $A_1$ and so on. A
reordering formula like \quince\ can also be obtained for
normal-ordered products of more than two fields. Proceeding as
we did to get eq. \quince\ and using the prescription
\dseis\ one obtains
$$
\eqalign{
(:ABC:)^n=&(-1)^{g(A)g(B)}(:BAC:)^n-(\partial D C)^n
-{\Delta_D\over 2}(\Delta_A-\Delta_B)k_D C^n\cr
(:CAB:)^n=&(-1)^{g(A)g(B)}(:CBA:)^n-(C\partial D )^n
-{\Delta_D\over 2}(\Delta_A-\Delta_B)C^n k_D,\cr}
\eqn\dsiete
$$
where we have supposed that the bracket \catorce\ still holds.

In order to define a topological theory we need to define a BRST symmetry.
We want this symmetry to be an odd
analogue of the current algebra. The best
way to proceed is by imitating the gauge-fixing procedure of Yang-Mills
theories. So, let us introduce a zero-dimensional anticommuting field
$\gamma_a$ and the following transformation:
$$
\delta J_a=if^{abc}\,\,\gamma_bJ_c-k\partial \gamma_a.
\eqn\docho
$$
It is easy to check that, if we want  eq. \docho\ to be compatible with the
commutation relations \tres, the constant $k$ must be the same in both
equations. We also have to specify what is the BRST transformation of the
ghost field $\gamma_a$. Requiring
nilpotency for the $\delta$-transformation,
one easily arrives at
$$
\delta\gamma_a={i\over 2}\,\,f^{abc}\,\,\,\gamma_b\gamma_c.
\eqn\dnueve
$$
Indeed, one can check using the Jacobi identity for the structure constants
of $g$,
$$
f^{abn}f^{ncm}+f^{bcn}f^{nam}+f^{can}f^{nbm}=0,
\eqn\veinte
$$
that
$$
\delta^2\gamma_a=\delta^2J_a=0.
\eqn\vuno
$$
We shall assign to $J_a$ and $\gamma_a$ ghost numbers $0$ and $+1$
respectively. From the transformation
laws (eqs. \docho\ and \dnueve) we see
that the $\delta$-variation increases in one unit the ghost number of the
fields.

The inclusion of the $\gamma_a$ field and the BRST transformations
(eqs. \docho\ and \dnueve) are not enough to define a topological field
theory.   In such a theory all the fields appearing in the  topological
algebra must have a   BRST partner. Therefore as we want to have a current
algebra  compatible  with the topological symmetry, it is clear that we must
introduce a new field such that its  $\delta$-variation is the {\bf total}
$g$-current of the theory. Let us denote it by  $\rho_a$.
As it is the partner
of the total $g$-current, it must be a dimension-one fermionic operator with
ghost number $-1$. With this ghost number and dimension, $\rho$ can be
naturally coupled to the  $\gamma$ field in the action by means of a term
$$
\int \rho_a \bar \partial \gamma_a
\eqn\vdos
$$
which implies that $\rho$ and $\gamma$ obey the following OPE
$$
\rho_a(z_1)\gamma_b(z_2)=-{\delta_{ab}\over z_1-z_2}
\eqn\vtres
$$
In terms of modes, eq. \vtres\ is equivalent to the following
anticommutators
$$
\{\rho_a^n,\gamma_b^m\}=-\delta_{ab}\delta_{n+m,0}.
\eqn\vcuatro
$$
Notice that the anticommuting fields  $\rho$ and $\gamma$ have color
indices. In fact they give a contribution to the total $g$-current, which
can be written as
$$
{\cal J}_a=J_a+if^{abc}\,\,\,:\gamma_b\rho_c:.
\eqn\vcinco
$$
It is straightforward to verify that one has
$$
\eqalign{
{\cal J}_a(z_1)\gamma_b(z_2)=&if^{abc}\,\,\,{\gamma_c(z_2)\over z_1-z_2}\cr
{\cal J}_a(z_1)\rho_b(z_2)=&if^{abc}\,\,\,{\rho_c(z_2)\over z_1-z_2}\cr}
\eqn\vseis
$$
which implies that, acting on
the fields $\gamma$ and $\rho$, ${\cal J}_a$ correctly generates non-abelian
rotations . As  was said above, we want $\rho$ to be the BRST partner of the
total current ${\cal J}$. Therefore we  write  $$
\delta\rho_a={\cal J}_a=J_a+if^{abc}:\gamma _b\rho_c:.
\eqn\vsiete
$$
Of course $\delta^2 \rho_a$  must vanish. By inspecting the right-hand side
of \vsiete\ we see that this requirement gives rise to a self-consistency
condition. In fact, performing a $\delta$-variation of the right-hand side of
\vsiete\  and using eqs. \docho\ and \dnueve, we get
$$
\delta^2\rho_a=-k\partial\gamma_a-{1\over
2}f^{abc}f^{brs}:(\gamma_r\gamma_s)\rho_c:
+f^{abc}f^{crs}:\gamma_b(\gamma_r\rho_s):,
\eqn\vocho
$$
where the parentheses in the last two terms indicate the normal-ordering
that results when the variation is performed. As  was mentioned above, the
normal order is not associative. Taking into account that
$$
[\gamma_b^n,(:\gamma_r\rho_s:)^m]=\delta_{sb}\gamma_r^{n+m},
\eqn\vnueve
$$
we can reorder the operators appearing in the last term of \vocho\ as
$$
:\gamma_b(\gamma_r\rho_s):=:(\gamma_r\rho_s)\gamma_b:-
\,\,\delta_{sb}\,\partial\gamma_r=-
:(\gamma_r\gamma_b)\rho_s:-\,\,\delta_{sb}\,\partial\gamma_r,
\eqn\treinta
$$
where eqs. \quince\ and \dsiete\ have been taken into account. Using eq.
\treinta\ in eq. \vocho, we simply get
$$
\delta ^2\rho_a=-(k+C_A)\partial\gamma_a,
\eqn\tuno
$$
\ie, only for the critical level $k=-C_A$ is our BRST symmetry nilpotent.
For this value of $k$ the total energy-momentum tensor becomes
$$
T=-{1\over C_A} :J_aJ_a:+:\rho_a\partial \gamma_a:,
\eqn\tdos
$$
where we have included the contribution of the $(\rho,\gamma)$ system.
It is easy to check that the operator $T$ in eq. \tdos\ has a vanishing
Virasoro anomaly. Indeed, particularizing eq. \trece\ to the case $k=-C_A$,
we obtain a value $2 \,{\hbox{\rm dim}}\,g$ for the central charge of the
bosonic $J_a$ currents. On the other hand, for fixed color index $a$, the
$(\rho_a,\gamma_a)$ fields form a spin-one $(b,c)$ system and therefore they
give a contribution $-2\,{\hbox{\rm dim}}\,g$ to the central charge, which
exactly cancels the one coming from the commuting sector of the theory. Thus
the model we have at hand is certainly topological. It is easy to see that
only for the critical value of $k$ determined above does the energy-momentum
tensor become  BRST-exact. Indeed, notice that the only field whose
$\delta$-variation produces the $J_a$ current is $\rho_a$. Therefore it is
clear that  by varying the operator $\rho_aJ_a$  we have a chance to
obtain the bosonic part of the energy-momentum tensor.  In fact we can
easily prove that
$$
\delta [ {1\over 2k+C_A}:\rho_aJ_a:]={1\over 2k+C_A}:J_aJ_a:+
{k\over 2k+C_A}:\rho_a\partial\gamma_a:,
\eqn\ttres
$$
where we have included a factor designed to reproduce the coefficient
appearing in the Sugawara expression (eq. \nueve). The right-hand
side of eq.
\ttres\ reproduces the energy-momentum tensor of the $(\rho,\gamma)$ system
if the coefficient of the $\rho\partial\gamma$ term is equal to one.
Again this only happens for the value $k=-C_A$. Therefore the BRST partner
of $T$ in eq. \tdos\ is
$$
G(z)=-{1\over C_A}:\rho_aJ_a:.
\eqn\tcuatro
$$

There is yet another argument to determine what level $k$  gives rise to a
topological theory. Consider the OPE of the total current ${\cal J}$ with
itself. A short calculation leads to
$$
{\cal J}_a(z_1){\cal J}_b(z_2)={(k+C_A)\delta_{ab}\over (z_1-z_2)^2}
+if^{abc}\,\,\,{{\cal J}_c(z_2)\over z_1-z_2},
\eqn\tcinco
$$
and, therefore,   $k=-C_A$ is the value for which the total Kac-Moody level
vanishes. Actually, for this value of $k$, ${\cal J}_a$ and its fermionic
counterpart $\rho_a$ close an algebra without any central extension:
$$
\eqalign{
{\cal J}_a(z_1){\cal J}_b(z_1)=&if^{abc}{{\cal J}_c(z_2)\over z_1-z_2}\cr
{\cal J}_a(z_1)\rho_b(z_2)=&if^{abc}{\rho_c(z_2)\over z_1-z_2}\cr
\rho_a(z_1)\rho_b(z_2)=&0.\cr}
\eqn\tseis
$$
{}From now on we shall put $k=-C_A$ in all our expressions.
Notice that, for an
abelian theory, $C_A$ vanishes and therefore $T$ and $G$ in eqs. \tdos\ and
\tcuatro\ become  ill-defined. In fact our non-abelian theory is only
consistent at the quantum level, since without taking  quantum
effects into account, we would have obtained the condition $k=0$, and the
classical energy-momentum tensor cannot be defined for such a value of $k$.

We shall call an algebra as the one displayed  in eq. \tseis\ a
{\bf topological current algebra}. These algebras, which
are the main subject
of this paper, are generated by two dimension-one operators, one bosonic
(${\cal J}_a$) and the other fermionic ($\rho_a$),
related by a topological
symmetry, in such a way that they form a BRST doublet. We now want  to
construct a topological algebra (containing $T$ and its partner $G$ among its
generators) compatible with the algebra \tseis. Let us see how this
construction can be carried out. First of all, let us write a local
expression for the generator of the topological BRST algebra:
$$
Q=-:\gamma_aJ_a:-{i\over 2}f^{abc}:\gamma_a\gamma_b\rho_c:.
\eqn\tsiete
$$
It can be checked that the $\delta$-variations \docho,
\dnueve\ and \vsiete\
are obtained by (anti)commuting with the zero-modes of $Q$. In fact the
OPE's of $Q$ with the fields of the theory are
$$
\eqalign{
Q(z_1)J_a(z_2)=&{1\over z_1-z_2}(if^{abc}\gamma_b(z_2)J_c(z_2)+C_A\partial
\gamma_a(z_2))+{C_A\over (z_1-z_2)^2}\gamma_a(z_2)\cr
Q(z_1)\gamma_a(z_2)=&{i\over 2}f^{abc}{\gamma_b(z_2)\gamma_c(z_2)\over
z_1-z_2}\cr
Q(z_1)\rho_a(z_2)=&{1\over
z_1-z_2}(J_a(z_2)+if^{abc}\gamma_b(z_2)\rho_c(z_2)),\cr}
\eqn\tocho
$$
while $Q$ acts on the ($\rho ,{\cal J}$) currents as
$$
\eqalign{
Q(z_1)\rho_a(z_2)=&{{\cal J}_a(z_2)\over z_1-z_2}\cr
Q(z_1){\cal J}_a(z_2)=&0.\cr}
\eqn\tochoi
$$
{}From the explicit expressions of $Q$ and $G$
(eqs. \tsiete\ and \tcuatro),
one can work out the OPE's of $Q$ with itself and
with $G$, with the result
$$
\eqalign{
Q(z_1)Q(z_2)=&0\cr
Q(z_1)G(z_2)=&{T(z_2)\over z_1-z_2}+{R(z_2)\over (z_1-z_2)^2}+{d\over
(z_1-z_2)^3}\, ,\cr}
\eqn\tnueve
$$
where $R$ and $d$ are, respectively,
a $U(1)$ current and a c-number given by
$$
R=:\rho_a\gamma_a:
\,\,\,\,\,\,\,\,\,\,\,\,\,\,\,\,\,\,\,\,
d= {\hbox{\rm dim}}\,\, g.
\eqn\cuarenta
$$
The first equation in \tnueve\ confirms that there are no anomalies that
could spoil the nilpotency of $Q$. The second one gives us how  $Q$ acts on
$G$. Notice that, acting with the non-zero modes of $Q$, the $R$ operator
and a c-number anomaly(proportional to the dimension of $g$) show up. The
interpretation of $R$ is quite clear: it is nothing but the ghost number
current. Indeed, one can check that all the fields transform under $R$ with
the ghost number we have assigned them. The second equation in \tnueve\ is
characteristic of the topological sigma models
\REF\wit{E. Witten \journal\cmp&118(88)411.}[\wit],
where the parameter $d$ is
the dimension of the target space in which the bosonic
sector of the theory is
embedded. For this reason $d$ is called the dimension of the topological
algebra. Notice that, in our case, $d$ is equal to the dimension of the
Lie group whose Lie algebra is $g$, which leads us to conclude that we
are describing a topological sigma model for the group manifold. After
all, this interpretation is quite natural, since what we are considering
is nothing but the Wess-Zumino-Witten model supplemented with a ghost
sector--- the level of the bosonic currents being chosen in such a way that
all local degrees of freedom can be gauged away. In order to establish the
nature of the topological symmetry, we have to compute all OPE's involving
$Q$, $R$, $G$ and $T$. After some calculations we get
$$
\eqalign{
T(z_1)Q(z_2)=&{Q(z_2)\over (z_1-z_2)^2}+{\partial Q(z_2)\over z_1-z_2}\cr
T(z_1)R(z_2)=&-{d\over (z_1-z_2)^3}+{R(z_2)\over (z_1-z_2)^2}+{\partial
R(z_2)\over z_1-z_2}\cr
T(z_1)G(z_2)=&{2G(z_2)\over (z_1-z_2)^2}+{\partial G(z_2)\over z_1-z_2}\cr
R(z_1)R(z_2)=&{d\over (z_1-z_2)^2}\cr
R(z_1)Q(z_2)=&{Q(z_2)\over z_1-z_2}\cr
R(z_1)G(z_2)=&-{G(z_2)\over z_1-z_2}.\cr}
\eqn\cuno
$$

The OPE's of eqs. \tnueve\ and \cuno\ are
also obtained when one performs a
twist to an $N=2$ superconformal algebra. However the $G(z_1)G(z_2)$ OPE
vanishes in  these algebras, while in our model we get
$$
G(z_1)G(z_2)={W(z_2)\over z_1-z_2},
\eqn\cdos
$$
where $W$ is a bosonic dimension-three operator
with ghost number $-2$, whose
explicit expression is
$$
W={i \over C_A^2}\,\,\,f^{abc}:J_a\rho_b\rho_c:-
{1\over C_A}\,\,\,:\partial \rho_a\rho_a:.
\eqn\ctres
$$
Remarkably, $W$ is $\delta$-exact. In order to check this fact, let us
consider the following operator:
$$
V={i\over 3C_A^2}\,\,\,f^{abc}:\rho_a\rho_b\rho_c:.
\eqn\ccuatro
$$
It is obvious that $V$ is an anticommuting
dimension-three field with ghost
number $-3$ and thus it is a good candidate to become the BRST partner of
$W$. Let us compute the OPE of $Q$ and $V$. From eq. \tochoi\ we get
$$
Q(z_1)V(z_2)={i\over C_A^2}{f^{abc}\over z_1-z_2}
:({\cal J}_a\rho_b\rho_c-\rho_a{\cal J}_b\rho_c
+\rho_a\rho_b{\cal J}_c):.
\eqn\ccinco
$$
Let us now reorder the last two terms in \ccinco\ by using our general
expressions \dsiete. Taking the OPE among the topological
currents (eq. \tseis) into account, we get
$$
\eqalign{
f^{abc}:\rho_a{\cal J}_b\rho_c:=&-f^{abc}:{\cal
J}_a\rho_b\rho_c:-iC_A:\partial\rho_a\rho_a:\cr
f^{abc}:\rho_a\rho_b{\cal J}_c:=&f^{abc}:{\cal
J}_a\rho_b\rho_c:+2iC_A:\partial\rho_a\rho_a:.\cr}
\eqn\cseis
$$
Substituting \cseis\ back into \ccinco\ and using the fact that
$$
f^{abc}f^{ade}:\gamma_d\rho_b\rho_c\rho_e:=0,
\eqn\csiete
$$
which is a consequence of the Jacobi identity (eq. \veinte), we can write
$$
Q(z_1)V(z_2)={W(z_2)\over z_1-z_2},
\eqn\cocho
$$
which proves our statement that $W$ is
$\delta$-exact. We must add these
new fields to the set of generators of our
topological algebra and we must compute their OPE's with any other
generator. There is no a priori guarantee that the algebra close within a
finite number of fields. However this is the case, as one can explicitly
check. One has:
$$
\eqalign{
Q(z_1)W(z_2)=&0\cr
R(z_1)W(z_2)=&-{2\over z_1-z_2}\,\,W(z_2)\cr
T(z_1)W(z_2)=&{2\over (z_1-z_2)^2}\,\,W(z_2)+
{\partial W(z_2)\over z_1-z_2}\cr
G(z_1)W(z_2)=&{2\over (z_1-z_2)^2}\,\,V(z_2)+
{\partial V(z_2)\over z_1-z_2}\cr
R(z_1)V(z_2)=&-{3\over z_1-z_2}\,\,V(z_2)\cr
T(z_1)V(z_2)=&{2\over (z_1-z_2)^2}\,\,V(z_2)+{\partial V(z_2)\over
z_1-z_2}.\cr}
\eqn\cnueve
$$
Other OPE's vanish, \ie
$$
G(z_1)V(z_2)=W(z_1)W(z_2)=V(z_1)W(z_2)=V(z_1)V(z_2)=0
\eqn\cincuenta
$$
Notice that, although our algebra involves dimension-three fields, it is
linear, contrary to what happens with the $W_3$ algebras.

We have thus found a topological algebra closed by three BRST doublets of
operators ($(Q,R)$, $(G,T)$ and $(V,W)$) of dimensions 1, 2 and 3. This
algebra is not unknown. It has been previously found by Kazama
\REF\kaza{Y. Kazama \journal\mpl&A6(91)1321.}[\kaza] as a
consistent non-trivial extension of the twisted $N=2$ superconformal algebra.
In Kazama's analysis no explicit representation of the algebra was given, and
it was related to an $N=1$ superconformal symmetry. A main outcome of
our representation is the connection between the extended character of the
algebra and non-abelian topological current algebras. In fact, apart from
those in eq. \tochoi, the only non-vanishing OPE's among the topological
currents $(\rho , {\cal J})$ and the generators of the extended algebra are
$$
\eqalign{
G(z_1){\cal J}_a(z_2)=&{\rho_a(z_2)\over (z_1-z_2)^2}+
{\partial \rho_a(z_2)\over z_1-z_2}\cr
G(z_1)\rho_a(z_2)=&0,\cr}
\eqn\ciuno
$$
which proves that, indeed, the topological and current algebras are
compatible. Notice that eq. \ciuno\ is the odd analogue of the first
equation in \doce. It is worth  pointing out that one can construct
an $N=2$
superconformal theory associated to the BRST charge $Q$ in eq. \tsiete\
\REF\Figueroa{J. M. Figueroa-O'Farrill, ``Affine algebras, $N=2$
superconformal algebras, and gauged WZNW models"(hep-th/9306164), Bonn
preprint BONN-HE-93-20.} [\Figueroa]. However, the energy-momentum tensor for
this theory does not have the canonical Sugawara form (eq. \tdos) and,
therefore, the currents are not primary fields.

An interesting feature of the realization we have obtained is that it can be
deformed in the following way:
$$
\eqalign{
T&\rightarrow T+\sum_a\alpha_a\partial {\cal J}_a\cr
G&\rightarrow G+\sum_a\alpha_a\partial \rho_a\cr
R&\rightarrow R+\sum_a\alpha_a {\cal J}_a,\cr}
\eqn\cidos
$$
where $\alpha_a$ are constant c-numbers.
The operators $Q$, $V$ and $W$ and the topological dimension $d$ are left
unaffected by the deformation. It can be easily verified that the
transformed operators also satisfy the extended topological algebra. Actually
eq.\cidos\ can be regarded as implementing a quantum Hamiltonian reduction
\REF\kpz{A.M. Polyakov \journal\mpl&A2(87)893; V. Knizhnik,
A.M. Polyakov and A.B. Zamolodchikov \journal\mpl&A3(88)819.}
\REF\BO{M. Bershadski and H. Ooguri \journal\cmp&126(89)49.}[\kpz,\BO] in
our theory. Of course when the transformation \cidos\ is  performed, the
currents are no longer primary dimension-one operators and, therefore, the
current algebra symmetry is lost.

Our construction closely resembles  the one performed in (super)string
theory, where the matter content of the theory is adjusted in such a way that
its total Virasoro anomaly exactly cancels the one coming from the
(super)conformal ghosts. This matching condition determines the critical
dimension of the theory. In our case  the condition required to
 have a topological current algebra is that the  total level $k$ of the
``matter" currents must equal $-C_A$, which is the contribution coming from
the ghost sector of the theory. Up to now, we have only considered the
situation in which only one bosonic current is present in the ``matter" part.
It turns out, however, that a similar construction can be performed with two
bosonic currents. Suppose we have two such independent currents $J_a^1$ and
$J_a^2$ that close an affine algebra, as in eq. \cuatro, with levels
$k_1$ and $k_2$ respectively. The total current is now
$$
{\cal J}_a=J_a^1+J_a^2 + if^{abc}\,\,\,:\gamma_b\rho_c:.
\eqn\citres
$$
If we assume that $J_a^1$ and $J_a^2$  transform under the
BRST symmetry as in eq. \docho\ and we impose   $\delta\rho_a={\cal
J}_a$ again, it is clear that we shall end up with the
following condition for the
two levels $k_1$ and $k_2$:
$$
k_1+k_2=-C_A
\eqn\cicuatro
$$
Observe that eq. \cicuatro\ only constrains the value of the sum of the
current algebra levels. Denoting $k_1$ simply by $k$ and using eq.
\cicuatro, we get for the energy-momentum tensor of the theory:
$$
T={1\over 2k+C_A} :J_a^1J_a^1:-\,\,{1\over 2k+C_A} :J_a^2J_a^2:
+:\rho_a\partial \gamma_a:.
\eqn\cicinco
$$
It is very easy to check that the operator $T$ in \cicinco\ has vanishing
Virasoro anomaly. The currents $J_a^1$ and $J_a^2$ give contributions
${2k\over 2k+C_A}\,{\hbox{\rm dim}}\,g$ and
${2k+2C_A\over 2k+C_A}\,{\hbox{\rm dim}}\,g$ respectively, whose sum exactly
cancels the central charge of the ghost system. Moreover it can be verified
that $T$ is also $\delta$-exact and, in fact, one has the same topological
algebra as in the one-current case. The generators are now given by
$$
\eqalign{
Q=&-:\gamma_a(J_a^1+J_a^2):-{i\over 2}f^{abc}:\gamma_a\gamma_b\rho_c:\cr
G=&{1\over2k+ C_A}:\rho_a(J_a^1-J_a^2):\cr
W=&{i \over (2k+C_A)^2}\,\,\,f^{abc}:(J_a^1+J_a^2)\rho_b\rho_c:-
{C_A\over (2k+ C_A)^2}\,\,\,:\partial \rho_a\rho_a:\cr
V=&{i\over 3(2k+C_A)^2}\,\,\,f^{abc}:\rho_a\rho_b\rho_c:.\cr}
\eqn\ciseis
$$
$R$ and $d$ are the same as in \cuarenta. Moreover the algebra is also
compatible with the current symmetry and eqs. \tochoi\ and \ciuno\ remain
valid; they are the only non-vanishing OPE's among the topological currents
and the generators of the extended algebra. This means that eq. \cidos\ also
provides a deformation of the theory that preserves its extended topological
symmetry. Notice that in this two-current case, our construction can be
performed for an abelian current algebra. However from eq.\ciseis\ we see
that $W$ and $V$ vanish identically in that case and we are left with the
standard $N=2$ twisted algebra (for $d=0$).

The form of $T$, $Q$ and $G$ in eqs. \cicinco\ and \ciseis\ is exactly the
one that appears in the $G/G$ coset models
\REF\witGG{E. Witten \journal\cmp&144(92)189.}
\REF\yank{M. Spiegelglas and S. Yankielowicz
\journal\np&393(93)301.}
\REF\aharo{O. Aharony et al. Tel Aviv University preprint,
TAUP-1961-92 \journal\pl&B289(92)309 \journal\pl&B305(93)35.}
\REF\hu{H.L. Hu and M. Yu \journal\pl&B289(92)302
\journal\np&B391(93)389.}[\witGG,\yank,\aharo,\hu]. These models
have a rich cohomology and, when they are deformed as in eq.\cidos, they
present striking similarities with minimal matter systems coupled to
two-dimensional gravity.

One might wonder if our construction can be carried out with an arbitrary
number of currents. If we had $M$ bosonic currents $J_a^i$ with levels $k_i$
($i=1,\cdots, M$), the nilpotency requirement imposes that $\sum_i
k_i=-C_A$. It is easy to verify that when $M \geq 3$ this condition for the
levels is not enough to ensure the vanishing of the Virasoro central charge.
Thus, in this case, by requiring the $\delta$-exactness of the total current
we do not eliminate all local degrees of freedom of the theory. We could, of
course, have $c=0$ by adjusting the $k_i$ levels without spoiling the
nilpotency of the BRST symmetry. However it is easy to see that $T$ cannot
be put as the $\delta$-variation of an operator local in the currents unless
$M$ is 1 or 2. Let us prove it. The Sugawara  energy-momentum tensor of
the theory is now
$$
T=\sum_{i=1}^M {1\over 2k_i+C_A}:J_a^iJ_a^i:+:\rho_a\partial\gamma_a:,
\eqn\cisiete
$$
The BRST-exactness of the total current implies that
$$
\delta \rho_a=\sum_{i=1}^MJ_a^i+ if^{abc}:\gamma_b\rho_c:.
\eqn\ciocho
$$
Let us study the possible form of $G$, the BRST partner of $T$. If we
require that $G$ be given by a local expression in the currents, it can only
have the form
$$
G=\rho_a\sum_{i=1}^M\lambda_iJ_a^i+Cf^{abc}:\rho_a\rho_b\gamma_c:,
\eqn\cinueve
$$
where $\lambda_i$ and $C$ are constants. The purely bosonic term in the
variation of $G$ can be easily computed
$$
\delta G |_{{\hbox {\sevenrm bosonic}}}=\sum_{i,j=i}^M\lambda_j:J_a^iJ_a^j:.
\eqn\sesenta
$$
Comparing this result with eq. \cisiete\ it is clear that we must require
that all crossed terms  in the double sum in \sesenta\ vanish,
\ie
$$
\lambda_i+\lambda_j=0
\,\,\,\,\,\,\,\,\,\,\,\,\,\,\,\,\,
{\hbox{\rm for}}\,\,\,\ i\not= j.
\eqn\suno
$$
Eq. \suno\ has   non-trivial solutions only if $M=1,2$, which are precisely
the ones we have written in eqs. \tcuatro\ and \ciseis.

To summarize, we have been able to formulate two-dimensional topological
conformal field theories possessing a current algebra symmetry. The currents
of these theories form a doublet under the topological symmetry and close
what we have called a topological current algebra, in which no central
extension appears. The price we have to pay for preserving the current
symmetry is a topological algebra involving dimension-three operators.
This makes a fundamental difference with the ordinary  Virasoro algebra, in
which the presence of a current symmetry does not require the modification
of the conformal algebra.

There are many directions in which our results can be generalized. We could,
for example, ask ourselves what happens if we want to make a supercurrent
algebra and a topological symmetry compatible. Presumably this is of
interest in the study of non-critical superstrings. Another topic that, in
our opinion, deserves further study is whether or not there is, in these
theories, an analogue of the coset construction, which would allow us to
generate many classes of topological matter. Finally, it might be possible
that some of the ideas developed here in a two-dimensional context could be
relevant in a three- or four-dimensional situation since, after all, gauge
symmetries exist in any number of space-time dimensions.

\ack
The authors would like to thank J.M.F. Labastida, J.~Mas, G. Sierra, L.
Alvarez-Gaum\'e and J.~S\'anchez Guill\'en for discussions. One of us (A.V.R.)
is grateful to the CERN Theory Division, where the last part of this work
was carried out, for hospitality. This
work was supported in part by DGICYT under grant PB90-0772, and by
CICYT under grants  AEN90-0035 and AEN93-0729.
\endpage

\refout
\end